# Nanotechnology and Innovation: Recent status and the strategic implication for the formation of high tech clusters in Greece, in between a global economic crisis

Evangelos I. Gkanas[1,a], Vasso Magkou-Kriticou[2,b], Sofoklis S. Makridis[1,c], Athanasios K. Stubos[3,d] and Ioannis Bakouros[4,e]

[1]*H2matters Group, Department of Mechanical Engineering, University of Western Macedonia, Bakola and Sialvera Street, 50100, Kozani, Greece*
[2]*Megalab SA, Stamati Psaltou 40, 54644, Thessaloniki, Greece*
[3]*IPTA, NCSR 'Demokritos', Agia Paraskevi, 13310, Athens, Greece*
[4]*Materlab Group, Department of Mechanical Engineering, University of Western Macedonia, Bakola and Sialvera Street, 50100, Kozani, Greece*
[a]*egkanas@uowm.gr,* [b]*info@megalab.gr,* [c]*smakridis@uowm.gr,* [d]*stubos@ipta.demokritos.gr,* [e]*ylb@uowm.gr*

## Abstract

*Nanotechnology is the first major worldwide research initiative of the 21st century and probably is the solution vector in the economic environment. Also, innovation is widely recognized as a key factor in the economic development of nations, and is essential for the competitiveness of the industrial firms as well. Policy and management of innovation are necessary in order to develop innovation and it involves processes. It is essential to develop new methods for nanotechnology development for better understanding of nanotechnology – based innovation. Nanotechnologies reveal commercialization processes, from start-ups to large firms in collaboration with public sector research. In the current paper, a study in the present status of innovation in nanotechnology and the affection of global economic crisis in this section is made and also the potential of increase the innovation via the presence of clusters in a small country like Greece which is in the eye of tornado from the global crisis is studied.*

## 1. Introduction

Nanotechnology has been defined as a multidisciplinary field in support of a broad – based technology to reach mass use by 2020, offering a new approach for education, innovation learning and governance [1]. Through is long – term planning, R&D (Research and Development) investments, partnerships, deliberate activities to promote public engagement anticipate the social consequence of scientific practices and integrate the social and physical sciences, nanotechnology is becoming a model for addressing the society implications and governance issues of emerging technologies generally [2]. The commercialized nanotechnology innovation that accomplices economic value for the nations that funded the research, requires a supportive investment and workface environment for manufacturing. Such environments has changed significantly in the last ten years by transfer of manufacturing capabilities from West to East and places risk in taking the nanotechnology benefits in the U.S.A and Europe as compared to Asia [3].

Managers of established companies wrestle with the discontinuous nature of an emergent technological trajectory [4], that changes the dynamics of intra – firm innovation processes [5]. Engineers and scientists are forced to navigate an interdisciplinary landscape of nanoscience and nanotechnology [6,7]. The nature of nanotechnology requires collaborative research efforts and nanotechnology innovation networks are therefore likely to be characterized by a degree of international and institutional diversity [8].

Huge public investments to support scientific and technological researchers [9] the creation of technological and industrial platforms and infrastructures [10] have led to more than 2.000.000 articles related with nanotechnology been published and over 1.000.000 applications lodged with patent office's [11]. Nanotechnology is something more than research as proved from the above. If this trend has good strategic and policy aims, then probably it will be helpful to humanity in many ways. Yet, a significant question remains: Is nanotechnology the next big thing in the scientific family which will revolutionize many industry sectors and will it bring radical change to may scientific and technological fields in ways that will benefit economies and consumers alike [12, 13]?

In the present study, we investigate the role of innovation evolution in nanotechnology in between a global economic crisis and we also try to find a reliable strategy for the increase the rate of innovation in Greece, which is a small country in southern Europe and nowadays has many economic problems and we also try to answer the following question: Will the increase of innovation in nanotechnology can help a country to become more competitive against the current situation?

In the following sections, we investigate the current status of nanotechnology worldwide and we try to understand the role of clusters of innovation, we explore the affection of the global economic crisis in nanotechnology research and finally we present the current situation of Greece in the innovation world map and try to explain how this situation can be improved and subsequently help country to avoid an economic breakdown.

## 2. Current Status of Nanotechnology and Future Predictions

Just a decade ago, governments, academia and industry, commissioned a massive expansion of research and development in nanotechnology – based on a long term science and engineering vision. Systematic investment in research on societal dimensions of nanotechnology has been undertaken in the United States since 2001, in the European Union since 2003, in Japan since 2006 and other countries since at least 2005. Nanotechnology has proven it has essential implications for how we comprehend nature, increase productivity, improve health and extend the limits of sustainable development among other vital topics.

Current nanotechnology developments have been successfully up to a point: products incorporating nanotechnology – based devices are on the markets, start – ups have been created and large firms have invested in production capacities [14, 15, 16]. Nanoscience and nanotechnology research is rapidly advancing, the rate of growth of the scientific production remains up to 10% per year, and nanotechnology based product innovations are increasing. Nanotechnologies are general purpose technologies [17]. This is the reason why they are the object of significant investments by incumbents [18].

There is also general agreement that nanotechnology is a platform technology with a potential to transform many industrial sectors in particular by fostering the convergence between previously separate technology – driven industries [19]. The interdisciplinary nature of nanotechnology that spans scientific developments across disciplines is also consistently highlighted [20]. The combination of newness and often asymmetric dispersion of knowledge about nanotechnology suggests that recent knowledge will most likely reside in networks of organizations, rather than individual members of a technology innovation system. [21]. Such networks can include individuals, firms, universities research institutes, venture capitalists and public policy agencies.

The key challenges to nanotechnology governance have been recognized and implemented. These, include developing the multidisciplinary knowledge foundation; establishing the innovation chain from discovery to societal use; establishing an innovative common language in nomenclature and patents; addressing broader implications for society, and developing the tools, people and organizations to responsibly take advantage of the benefits of this new technology. To address these challenges, there are four simultaneously characteristics of effective nanotechnology governance which proposed and applied since 2001. Nanotechnology governance needs to be: Transformative, which includes the results of project – oriented focus on advancing multidisciplinary and multisector innovation , Responsible, which includes EHS and aquitable access and benefits, Inclusive and Visionary, which includes long – term planning and anticipatory adaptive mesurments.

Overall, the governance of nanotechnology has been focused on the first generation of nanotechnology products (passive nanostructure) with the main research and studies commencing on the next generations.

## 3. Nanotechnology: Innovation and Commercialization

Innovation in nanotechnology, generally involves a complex value chain, including large and small companies, research organizations, equipment suppliers, intermediaries, finance and insurance, end users (who may be in the private and public sectors), regulators and other stakeholder groups in a highly distributed global economy [22, 23, 24, 25]. Between 1990 and 2008 about 17600 companies worldwide published about 52600 scientific articles and applied for 45.052 patents in the nanotechnology domain. The ratio of corporate nanotechnology patent applications to corporate nanotechnology applications increased rapidly from about 0.23 in 1998 to 1.2 in 2008.

Due to the nature of nanotechnology, indicates that many geographical regions will have opportunities to engage in the development of nanotechnology. For example while leading high technology regions in the United States are at the forefront of nanotechnology innovation, some other U.S cities and regions also have clusters of corporation engaged in nanotechnology innovations.

A key factor for commercialized innovation and economic development is the nanotechnology development and the «general technology development strength» of each nation [26]. Some other key factors for innovation and corporate decision making in nanotechnology are recognizing consumer values, their perception of acceptability of the products and their response to labeling. Consumer perception are affected by awareness education and access to information. According to D.A. Hart [27] there are four explicit, and one implicit, aspects of the definition of innovation are important to us. In terms of the explicit dimensions:

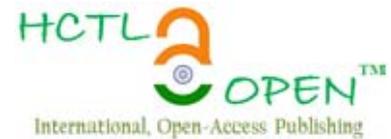

Firstly, innovation is a commercial concept not simply a technological, or even an intellectual property one. However novel an innovation is, unless firms are able to successfully exploit their innovation in commercial terms it is not relevant for our present purposes.

Secondly, there are degrees of innovation. The innovative process can involve the creation of completely new products or services or, more commonly, simply the improvement of existing products and services. Innovation can thus be radical or incremental in character.

Thirdly, whatever the degree of innovation it normally arises because individuals working in groups have learned from each other how new or improved goods and services can be created and commercially exploited.

Fourthly, the basic unit of innovative process is not necessarily an individual, or even an individual firm working in isolation, it is a network of individuals, or firms, working together to produce the innovation.

As noted in the above the role of nanotechnology in increasing the innovation process is very important and significant. Also the nature of nanotechnology gives the chance to small countries and regions to develop the own regional innovation. The best strategy for doing these is the establishing of innovation clusters.

## 4. Characterization of regional clusters

Nations and regions are struggling to remain competitive and adapt in the context of globalization. The regional specializations built up over decades are transforming rapidly. Many regions that were historically production centers are losing out to lower-cost locations and are reorienting their activities to higher value-added non-manufacturing industries or R&D-intensive manufacturing niches.

 The public sector response has been an increased attention to the importance of linking firms, people and knowledge at a regional level as a way of making regions more innovative and competitive. This new approach is visible across a number of different policy fields. Evolutions in regional policy, science and technology policy and industrial/enterprise policy are converging on the objective of supporting these linkages at the regional level. One of the vehicles commonly used to achieve these goals is to support "clusters" (concentrations of firms and supporting actors) in a particular region. Examples of such programs include the Pôles de Compétitivité in France, the Centers of Expertise in Finland or Japan's Industrial Clusters [28].

"Regional clustering" has been used to describe industrial districts of small crafts firms, high technology centers, agglomerations of financial and business service firms in cities, company towns, and large branch plants and their supply chains [28] The geographic scope of a cluster refers to the territorial extent of the firms, customers, suppliers, support services, and institutions that are embedded in the ongoing relationships and interdependent activities that characterize the cluster. The geographic span of a cluster can range from a small area within a city to areas encompassing much of a nation. [29].

The breadth of clusters refers to the range of horizontally related industries (industries related by common technologies, end users, distribution channels, and other non-vertical relationships) within the cluster. Narrow clusters consist of one of a few industries and their supply chains. Broad clusters provide a variety of products in closely related industries [30]. The activity base of a cluster involves the number and nature of the activities in the value added chain that are performed with the region. In activity-rich clusters, most or at least many of the critical activities in the value-added chains of the relevant industries are performed locally. Firms in such clusters tend to carry out the core strategy-setting, product or service development, marketing strategy, and corporate co-ordination activities within the region in question. Activity-poor clusters, on the other hand, involve one or only a few activities in a given industry or set of related industries.

The innovative capacity of the cluster refers to the ability of the cluster to generate the key innovations in products, processes, designs, marketing, logistics, and management that are relevant to competitive advantage in the industries in question. The distinction between high innovation and low innovation clusters is far more useful than that between "high technology" and "low technology" clusters. Some "high technology" industries are not at all innovative and some "low technology" clusters are. A cluster's ability to sustain itself is related more to its innovative capacity than to the level of technology produced or used in the process [30]

## 4.1. Analyzing the various strategies to support regional specialization and clusters reveals the following trends

Regional policy: capitalizing on local assets. Cluster policies linked to regional policy usually focus on so-called lagging regions, including regions undergoing industrial restructuring and geographically peripheral regions. Such programs often use EU Structural Funds. In addition, several initiatives originating in other policy families have incorporated a clear regional dimension, indicating the recent emphasis in science and technology and enterprise policy on regions, such as regional innovation system concepts.

S&T and innovation policy: from research to economic growth. Several of the more recent cluster/regional specialization programs were born from science and technology policy. They promote collaborative R&D to support growth of the most promising technology sectors in regions where key institutions, researchers and firms are concentrated.

Industrial and enterprise policies: supporting groups not firms. Industrial policies with cluster programs tend to support those clusters that drive national growth, with business linkages taking priority over research initiatives. These trends illustrate an evolution from prior industrial policies to support strategic sectors and work with individual large firms. The cluster approach provides a more transparent and less trade-distorting framework for efforts to strengthen strategic sectors. Programs that originate from an enterprise policy tend to focus more on SME clusters. These include a number of programs started as early as the 1980s that emphasize the industrial district model of cluster policies. Programs that focus on disadvantaged regions also tend to be closely linked with SME policy, emphasizing the widely-held policy objective of building critical mass (for export, for access to information, etc.) among SMEs.

Linking objectives and changing objectives. It is more common than not that policies to promote clusters link multiple objectives. Furthermore, the objectives of these programs appear to change over time within a given country depending on economic needs and changes in the popularity of the

Policy approach. Over time, these policies have generally transitioned from SME-based programs to those supporting national competitiveness clusters and they increasingly focus on technology and innovation.

Table 1. Policy trends supporting clusters and regional innovation systems[31]

| Policy stream | Old approach | New approach | Cluster programme focus |
|---|---|---|---|
| Regional policy | Redistribution from leading to lagging regions | Building competitive regions by bringing local actors and assets together | Target or often include lagging regions<br>Focus on smaller firms as opposed to larger firms, if not explicitly than *de facto*<br>Broad approach to sector and innovation targets<br>Emphasis on engagement of actors |
| Science and technology policy | Financing of individual, single sector projects in basic research | Financing of collaborative research involving networks with industry and links with commercialization | Usually a high-technology focus<br>Both take advantage of and reinforce the spatial impacts of R&D investment<br>Promote collaborative |
| | | | R&D instruments to support commercialization Include both large and small firms; can emphasize support for spin-offs and start-ups |
| **Industrial and enterprise policy** | Subsidies to firms; national champions | Supporting common needs of firm groups and technology absorption (especially SMEs) | Target the drivers of national growth<br>• Support industries undergoing transition and shedding jobs<br>• Help small firms overcome obstacles to technology absorption and growth<br>• Create competitive advantages to attract inward investment and branding for exports |

Some groups [27, 30] characterize clusters by their "type" or the extend in which the cluster exists:

Working clusters are those in which a critical mass of local knowledge, expertise, personnel, and resources create agglomeration economies that are used by firms to their advantage in competing with those outside the cluster. Working clusters tend to have dense patterns of interactions among local firms that differ quantitatively and qualitatively from the interactions that the firms have with those not located in the cluster. They often have complex patterns of competition and co-operation and often are able to attract mobile resources and key personnel from other locations. Even if participants do not call themselves a "cluster" there tends to be knowledge of the interdependence of local competitors, suppliers, customers, and institutions.

Latent clusters have a critical mass of firms in related industries sufficient to reap the benefits of clustering, but have not developed the level of interaction and information flows necessary to truly benefit from co-location. This can be due to a lack of knowledge of other local firms, a lack of interaction among firms and individuals, a lack of a common enough vision of their future, or a lack of the requisite level of trust for firms to find and exploit common interests. In any case, such groups of firms do not think of themselves as a cluster and, as a result, do not think of exploring the potential benefits of closer relationships with other local organizations.

Potential clusters are those that have some of the elements necessary for the development of successful clusters, but where these elements must be deepened and broadened in order to benefit from the impact of agglomeration. Often there are important gaps in the inputs, services, or information flows that support cluster development. Like latent clusters, they lack the interaction and self-awareness of working clusters.

Policy driven clusters are those chosen by governments for support, but which lack a critical mass of firms or favorable conditions for organic development. Many of the electronics and biotechnology "clusters" found in government programs are examples of this type of cluster. Policy driven clusters tend to be chosen more on political grounds than through any detailed analytical process. They tend to rely on the notion that policy can create clusters from a relatively unfavorable base.

"Wishful thinking" clusters are policy driven clusters that lack, not only a critical mass, but any particular source of advantage than might promote organic development.

Although preliminary in nature, such results should prove very useful to researchers investigating the regional clustering phenomenon as well as development professionals and policy makers interested in formulating and executing development policies based on regional clustering.

## 6. Crisis, recovery and the role of innovation

The European Union is recovering from the effect of the major global crisis in 2008-2010. The recession originated from the accumulation of considerable imbalances in the pre-crisis period 2000-07, notably the inflation of house and stock prices in the US and some EU Member States, and the subsequent unbalanced capital flows. [32]

The crisis has affected all EU Member States and, with the exception of Poland and Slovakia, no country experienced less than a full year of recession. Even if by mid-2009 most countries had started to recover, some Member States like Greece, Ireland or Romania were still in recession by the beginning of 2011: after almost three consecutive years of decreasing income. The experience is also mixed when it comes to the depth of the recession, ranging from a tiny one-quarter point drop in Poland to a 25 percent loss during the more than two years of recession in Latvia. The reason is that not all countries played the same role during the accumulation of these imbalances and, consequently, not all countries are affected in the same way. On the one hand, countries like Latvia, Ireland or Spain, which were severely affected by a housing bubble, are now going through a major readjustment. On the other hand, there are countries like Austria, Belgium or Germany that can be seen mostly as suffering the collateral effects from the readjustments in the US and in the first group of Member States; these countries have been affected chiefly through international trade, but also through the exposure of their financial systems to loans made to countries with large imbalances [33].

The focus here is on R&D and innovation, because it is regarded as an important source of sustained growth. The EU is characterized by a lower intensity than the US and a remarkable heterogeneity in R&D intensity across Member States. However, a closer look at the individual US states shows that the internal variability is not different from that within the EU. This variability reflects patterns of regional specialization which may be optimal from the social point of view. In that sense, it is worth recalling that the new Europe 2020 strategy maintains the Lisbon strategy target of a 3 percent for R&D intensity for the EU as a whole (rather than for each individual Member State) [35].

One possible explanation for these differences is that EU Member States tend to specialize in sectors characterized by a lower R&D intensity. However, a closer look at the figures shows that, even if the sectorial composition plays a role, most of the differences with the US can be associated with lower EU intensities in individual sectors rather than an over-representation of low-intensive sectors in the EU. Furthermore, when comparing similar firms from across the Atlantic, they turn out to be remarkably similar in that they are making similar efforts in terms of R&D. These two pieces of evidence together show the frequency with which we find innovative firms in the US being compared with the EU. Hence, the key area is the relatively poor commercialization of R&D and non-technological innovation in the EU, rather than R&D per se. The EU must therefore do more than just foster basic research in order to create ideas, and it needs to create the right business conditions for new technologies and innovations to be developed and commercialized on the market. The whole process has to be complemented by an adequate level of intellectual property rights protection: enough to give incentives to innovators but not so much that it hampers the creation of new ideas or shifts research too soon away from academia by offering excessive incentives to privatize basic lines of research. The EU is currently working with a High Level Group of experts to examine how to improve the commercialization of key enabling technologies [34].

## 7. Innovation activities in Greece.

Some researchers [35] have been performed a research to Greek innovation in manufacturing enterprises in Greece until 2000. The results of the research are seen in Table 2.

Table 2. Innovation in manufacturing enterprises in Greece.

| Indexes | 1994-96* | | 1996-98* | | 1998-00** | |
|---|---|---|---|---|---|---|
| | % share in population | % innovative enterprises | % share in population | % innovative enterprises | % share in population | % innovative enterprises |
| Enterprises with innovation activity | 26,50 | 100 | 30.3 | 100 | 27.3 | 100 |
| Product innovators | 22,5 | 85.1 | 25.2 | 83.3 | 18.4 | 67.3 |
| Process innovators | 18,5 | 70.2 | 23.7 | 78.1 | 17.5 | 64.1 |
| Intramural R&D | 20,6 | 77.9 | 21.2 | 69.8 | 21.8 | 79.8 |
| Research and experimental development - R&D | 15,8 | 59.7 | 18.9 | 62.3 | 17.3 | 64.7 |
| Continuous R&D | 5,1 | 19.3 | 7.1 | 23.3 | 7.1 | 26.1 |
| Occasional R&D | 10,7 | 40.3 | 11.8 | 39.1 | na | na |
| Enterprises with Cooperation arrangements on innovation activities | 4,7 | 17.7 | 6.5 | 21.4 | 5.1 | 19.9 |
| Product innovators that introduced | 10,4 | 39.2 | 14 | 46 | 10.3 | 37.8 |

| | | | | | |
|---|---|---|---|---|---|
| **new or improved products to the market** | | | | | |
| **Enterprises receiving public funding** | 11,4 | 43.1 | 10.9 | 35.8 | 17.0*** | 16.4 |

Sources: GSRT (2001), European Commission, (2004),
Note:*>20 employees;**>10 employees;***Central government

According to them, only a small part of Greek firms has received financial support either from local – regional authorities or from central government and European Union. In particular, a a part accounting 18 % of Greek firms have received financial support for innovation activities through European Union, while the other sources of financial support, namely the central government and the local-regional authorities, account around to 6 % and 3 %, respectively. Furthermore, the main obstacles in relation to «economic factors» the lack of appropriate financial sources for innovation activities that accounts 70 % for small firms and only 10 % for big firms, while the high risk activities accounts around to 74 % for small firms and only 5 % for big firms, respectively. Finally, the main obstacles in relation to the «internal factors» are the lack of information of new technologies and for the markets accounting 75 % for small firms and 25 % for medium firms, while the lack of specialized staff accounts 60 % for small firms and 40 % for medium firms, and the managerial inflexibilities that accounts 40 % for the small firms and 40 % for the medium firms, respectively.

Comparisons [36] show that Greece performance is generally lagging in most dimensions of innovation with rankings significantly lower than those achieved by other small countries that are leaders in innovation. This makes the task of improving Greece's performance especially daunting given that reforms will have to be simultaneously implemented across many policy areas and levels. In the majority of the indicators, the rankings of Greece are below the average ranking of the European Union (EU). In particular the country's ranking in R&D expenditures, in firms capacity to innovate, and in trademarks and patents is especially low. Other areas with significant underperformance appear to be found in the quality of the educational system, in the University-Industry relationships, in business start-up requirements, and in technology infrastructure.

Specifically, Greece appears to be particularly open to new ideas (as per the indicator on "national culture adaptation to new ideas") but underperforms in the final result, the implementation of the ideas. Total business expenditures on innovation (in a wider comprehensive sense, including expenditures beyond R&D) as well as public subsidies for innovation are also high. However they do not seem to drive the country to higher innovation rankings. Tertiary educational attainment is high and so is the availability of scientific personnel. But achievement on these dimensions may be offset, at least to an extent, by the low quality of the educational system, as suggested by the respective indicators.

In comparison with other countries, Greece does not lack in innovation policies and programs. Actually, in many cases (e.g. innovation programs) Greece applies policies that are European best practices. However, the plethora of policies and programs has not been followed by results. National resources are dispersed into many programs that cover all aspects of the modernization of business enterprises, with a low and rather vague threshold requirement in order to be categorized as innovative. [36, 37]

## 8. Clustering Greek Regional Projects

In this section we investigate the point in which Greece lies as a member of E.U in clustering projects. Cluster policy is a 'mature' policy area in some countries, and one that is emerging in others. Denmark was among the pacesetters in developing cluster policies with its Industrial Network Co-operation Program. Other successful examples of clusters are the Italian Industrial Districts, the French Systemes Productifs Locaux, the British Business Networks and the Finnish Centers of Excellence. A good cluster is like a "sponge" - it can absorb and retain knowledge, skills and activity. The question for regions and governments is how they can cultivate such "sponges"

The following table 3 displays some EU countries according to their cluster policy type [23].

Table 3. Classification of European Union countries according to their cluster policy type.

| National Policy | France, Luxembourg, Latvia, Lithuania, Slovenia |
|---|---|
| Regional Policies | Belgium (Wallonia, Flanders and Brussels regions), Spain |
| National Frameworks for regional policies | Austria, Germany, Hungary, Italy, Sweden, UK |
| Scarce Policy attempts | Czech Republic, Estonia, Denmark, **Greece**, Ireland, Netherlands, Poland, Portugal, Slovak Republic |

The characteristic of the Greek economy (a small-medium size economy) revealed that only a small number of industries and clusters are present. To build up clusters, in a pragmatic way, we have to begin with small groups of obviously related industries and subsequently discover further correlation patterns. The regional economic activity by industry can be broken down as follows:

• In-region oriented (Local) - Local industries provide goods and services almost exclusively for the area in which they are located.

• Out-region oriented (Traded) - Traded industries sell products and services across regions and frequently to other countries. They are located in a particular region not because of the available natural regional resources or regional selling potential but due to broader location-based competitive advantages. According to their stage of development, all Greek clusters are classified as embryonic. Based on their depth, diversity and range of industries that could be found present within an identified cluster, they are characterized as shallow.

Based on an assessment of their significance, Greek clusters could be classified as being of national importance but as having limited potential for achieving international significance in a couple of sectors (i.e. tourism).

There are also a number of "unique" clusters, mainly linked to industries that have developed around regional natural resources (i.e. electricity, coal mining).

It should be noted that a study entitled "The Future of Greek Industry", commissioned in 1997 by the Ministry of Development, demonstrated the existence of networking in industries (i.e. furniture, solar energy panels, wine, food, marble, tourism, fur, software) that could be upgraded to potential clusters.

According to the fifth edition of the EIS (European Innovation Scoreboard) [European Innovation Scoreboard 2008 Comparative Analysis of innovation performance - 2009] the overall innovation performance classification of the European countries can be grouped in four clusters:

1)  Sweden, Finland, Germany, Denmark and the UK are the Innovation leaders, with innovation performance well above that of the EU average and all other countries.

Of these countries, Germany is improving its performance fastest while Denmark is stagnating.

2)  Austria, Ireland, Luxembourg, Belgium, France and the Netherlands are the Innovation followers, with innovation performance below those of the innovation leaders but above that the EU average. Ireland's performance has been increasing fastest within this group, followed by Austria.

3)  Cyprus, Estonia, Slovenia, Czech Republic, Spain, Portugal, Greece and Italy are the Moderate innovators, with innovation performance below the EU average. The trend in Cyprus' innovation performance is well above the average for this group, followed by Portugal, while Spain and Italy are not improving their relative position.

4)      Malta, Hungary, Slovakia, Poland, Lithuania, Romania, Latvia and Bulgaria are the Catching-up countries with innovation performance well below the EU average. All of these countries have been catching up, with the exception of Lithuania. Bulgaria and Romania have been improving their performance the fastest.

In Figure 1 shows the summary innovation performance of E.U members states for 2009

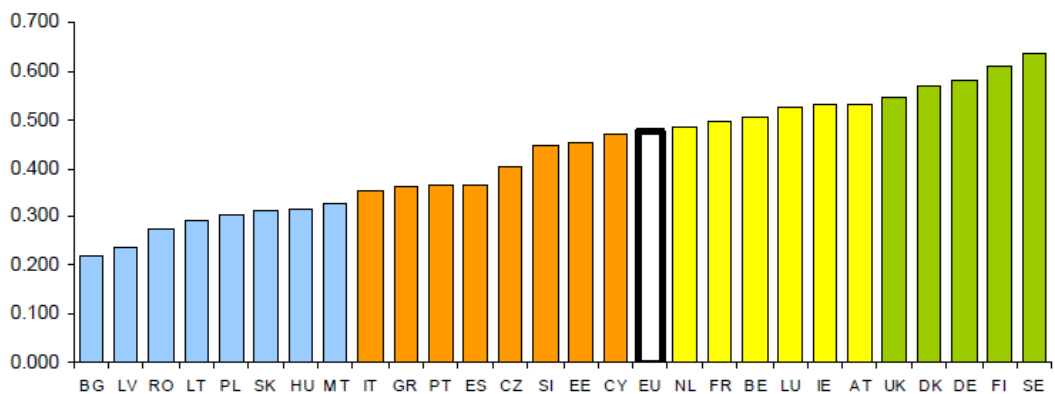

Fig. 1. Summary innovation performance of E.U members states for 2009 [38].

In particular, the thirteen Greek regions are part of a team of 56 regions characterized by (1) lower employment level in hi-tech, (2) lower business R&D expenditure, (3) almost null patent records and (4) lower educational level.

Table 3. Distribution of geographical regions according some characteristics of innovation [23].

| | Cluster 6 | Cluster 5 | Cluster 4 | Cluster 3 | High tech cluster 1 | High tech cluster 2 | Total number of regions |
|---|---|---|---|---|---|---|---|
| **Regions** | **56** | **65** | **28** | **10** | **3** | **3** | **1711** |
| **Austria** | 1 | 8 | | | | | 9 |
| **Belgium** | | 2 | 1 | | | | 3 |
| **Germany** | | 28 | | 10 | | 2 | 40 |
| **Greece** | 13 | | | | | | 13 |
| **Spain** | 12 | 3 | 2 | 1 | 1 | | 171 |
| **Finland** | | 1 | 3 | 2 | | | 6 |
| **France** | 9 | 11 | | | | | 221 |
| **Ireland** | | 2 | | | | | 2 |
| **Italy** | 14 | 6 | | 1 | | | 20 |
| **Netherlands** | | 4 | 6 | | | 1 | 12 |
| **Portugal** | 7 | | | 2 | 2 | | 7 |
| **Sweden** | | | 4 | | | | 8 |
| **U.K** | | 12 | | | | | 12 |

According to INNO Metrics website [38], Greece is one of the moderate innovators with a below average performance. Relative strengths are in Human resources, Linkages & entrepreneurship and Innovators. Relative weaknesses are in Finance and support, Firm investments and Intellectual assets. These are shown in Figure 2.

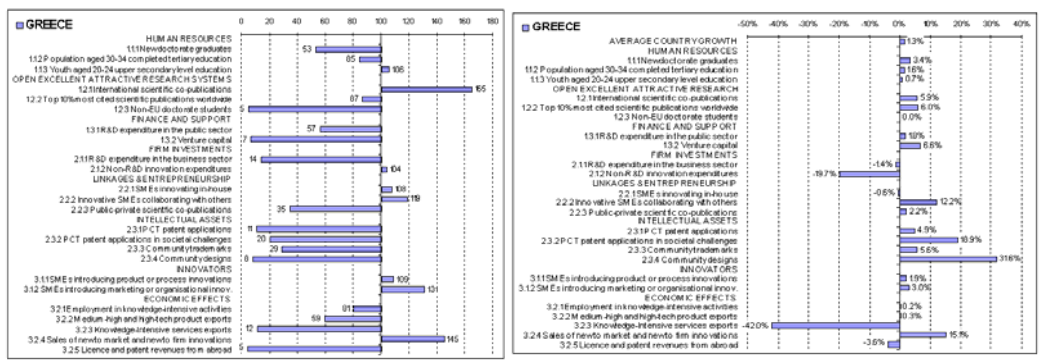

Figure 2. Indicator values relative to the EU27 (EU27=100) (left) and Annual average growth per indicator and average country growth (right) [38].

Figure 3 represents the classification over the European countries according the number of employees in clusters. Russia, Germany, Italy, France and Spain seem to have the biggest potential, while Greece is very weak in this direction compared with the neighbor countries.

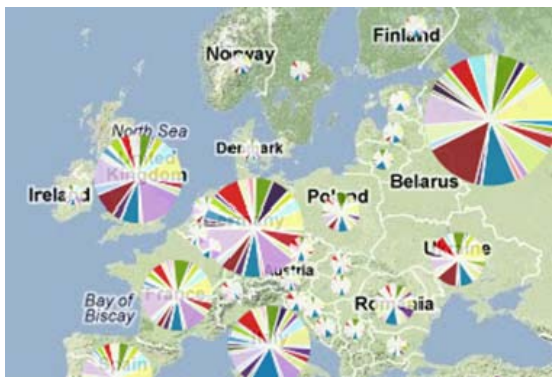

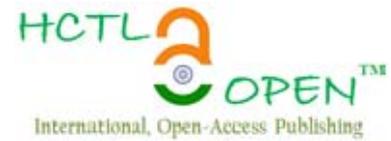

Figure 3. Cluster indicator according to employees over Europe Source: http://www.clusterobservatory.eu

As a result from the above analysis, Greece is a moderate innovator, with innovation performance below the EU average. Also every region Greece is characterized by the lower employment level in hi-tech, the lower business R&D expenditure, almost null patent records and lower educational level.

## 9. Conclusions

Exploring a Greek model of innovation may entail putting emphasis on the adoption and adaptation of proven technologies and solutions through small – incremental innovations, applications in new context, in their adaptation to consumer needs, in customer service, or in internal organizational processes. This is probably more operational in an economy wide scale than emphasizing a model focused on basic, radical innovations. Such incremental adaptations and improvements may be inspired and enriched by the Greek reality, the rich traditions and social values.

Small countries, like Greece, is likely to need a more comprehensive and oriented policy of co-operative innovative effort, in order to develop their future capabilities and to make the necessary choice for technological priorities. The various most important factors which might influence the incidence of innovation and the speed of its diffusion are technical applicability, profitability, finance, (lack of financial resources might delay the diffusion of new processes, size - structure and organization, management attitudes (which is the most difficult to assess or to quantify, but nevertheless they may be as important as economic factors in influencing the rate of adoption of new methods), and finally other factors, such as research and development activities, access to information, the labor market availability of certain skills, licensing policy, the market situation and more precisely the growth of demand for the product as well as the competitive position with special regard to the import competition. All these illustrate the wide range of factors which could contribute to explain the differences in the speed of diffusion.